# Real-time Detection and Auto focusing of Beam Profiles from Silicon Photonics Gratings using YOLO model

Yu Dian Lim[1,*], Hong Yu Li[2], Simon Chun Kiat Goh[3], Xiangyu Wang[2], Peng Zhao[1,^], and Chuan Seng Tan[1,2]

[1]*School of Electrical and Electronics Engineering, Nanyang Technological University, 639798, Singapore*
[2]*Institute of Microelectronics, Agency for Science, Technology and Research (A*STAR), 117685, Singapore*
[3]*Device Solutions Research Singapore, Samsung Electronics, 117440, Singapore*
*^current address: imec, Leuven, 3000, Belgium*
*\* yudian.lim@ntu.edu.sg*

**When observing the chip-to-free-space light beams from silicon photonics (SiPh) to free-space, manual adjustment of camera lens is often required to obtain a focused image of the light beams. In this letter, we demonstrated an auto-focusing system based on you-only-look-once (YOLO) model. The trained YOLO model exhibits high classification accuracy of 99.7% and high confidence level >0.95 when detecting light beams from SiPh gratings. A video demonstration of real-time light beam detection, real-time computation of beam width, and auto focusing of light beams are also included.**

Over the past decade, silicon photonics (SiPh) has attracted much attention among both research communities and industrial players. As the world enters a new era of artificial intelligence (AI), with a growing demand for computing power and advanced algorithms, energy consumption is becoming an increasingly significant concern. Attributing to its low-energy-consumption nature, SiPh integrations are expected to reduce the energy consumption from massive computing activities[1]. In the context of reducing the energy consumption for AI computing, various SiPh devices, including modulators[2], III-V/Si lasers[3], and optical frequency comb[4] have been developed in wafer-scale for the electrical/optical signal switching[5]. Apart from conventional AI computing, the applications of integrated SiPh have been explored for quantum computing applications. In the previously-reported works [6,7], SiPh waveguide/grating structures are buried underneath the planar electrode of ion trap in an integrated ion trap chip. Then, laser light of specified wavelengths is coupled into the chip, propagates through the waveguide, and coupled out from grating into free space. The light in free-space shall reach the trapped ion qubits to optically-address them, for their applications in quantum computing operations. The wavelengths of laser light depend on the atomic-energy level of the trapped ions[8]. The abovementioned optical addressing technique is illustrated in Fig. 1.

As shown in Fig. 1, the chip-to-free-space light from gratings propagates upwards in a typical Gaussian beam behavior[9], where it focuses at a specific height along z-axis, and dispersed as it propagates further from the focal point. For optimized optical addressing of ions, trapping the ions at the focal point of the beam is crucial. At the same time, thorough investigations on the beam characteristics, including the beam width and beam morphologies is crucial for development of SiPh gratings for the optical addressing of trapped ion qubits.

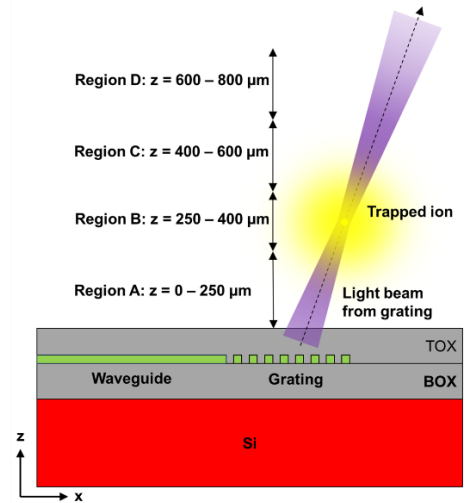

Fig. 1. Illustration of optical addressing using SiPh gratings, and categorization of beam images in Region A, B, C and D.

To facilitate this investigation, in this letter, we proposed a backend system that performs auto-focusing of CMOS camera when viewing the light beam coupled out from SiPh gratings to free-space. The system is based on the state-of-the-art You-Only-Look-Once, version 8 (YOLOv8) model. Benchmarking against the state-of-the-art, Hu *et al.* developed a continuous geometric model to achieve auto-focusing 3D measurement[10]. Meanwhile, Zhou *et al.* developed an autofocusing algorithm using pixel difference with the Tanimoto coefficient (PDTC) to predict the focus[11]. However, there has been limited work on implementing YOLO model for auto-focusing of light beams, which can detect, focus, and compute the beam width of the light, continuously.

To summarize the concept of the YOLO-based auto-focusing system, the trained YOLO model will first perform real-time detection when viewing the light beams from SiPh gratings. When unfocused beams are detected, the camera will move automatically to reach the position where focused beams can be seen. At the same time, real-time computation of beam width has also been included in the system. The demonstration video of the real-time detection and auto-focusing has been uploaded to YouTube [12].

**YOLO Model Training.** The SiPh grating structure used in this work is shown in Fig. 2(a), and the microscopic images of the grating were taken using the camera system shown in Fig. 2(b). The wavelength of the laser light used in this work is 1,092 nm, which corresponds to the 'clear-out' function of $^{88}Sr^+$ trapped ion qubit [13]. As the 1,092 nm is near the visible region of 400 – 700 nm, the light beams from the output gratings are visible under the CMOS camera shown in Fig. 2(b). The gratings are 0.4 µm-thick silicon nitride etch-through gratings. The input grating has fixed pitch of 0.8 µm, while the pitches of the output gratings varied from 0.6 to 1.2 µm to cater for output light beams with various intensities. The duty cycle of the gratings is fixed at 0.5, and the dimension of the gratings are fixed at ~55×55 µm.

To develop the real-time detection and auto-focusing system, first, the distance between the camera lens and the sample chip is adjusted to obtain the sharpest possible image of the grating structures on the camera. At this position, the light beam images taken are defined as z = 0 µm. After that, the z-axis motorized stage continuously moves the camera lens upwards (along z-axis). For every 1 µm elevation, an image is taken. The elevation-image capturing cycle is repeated from z = 0 to 800 µm. The abovementioned process is repeated for 4 times with different laser intensities and focusing region. A total of 3,200 microscopic images are taken for the training of YOLO model. Each image is fixed to have 10 light beams in it.

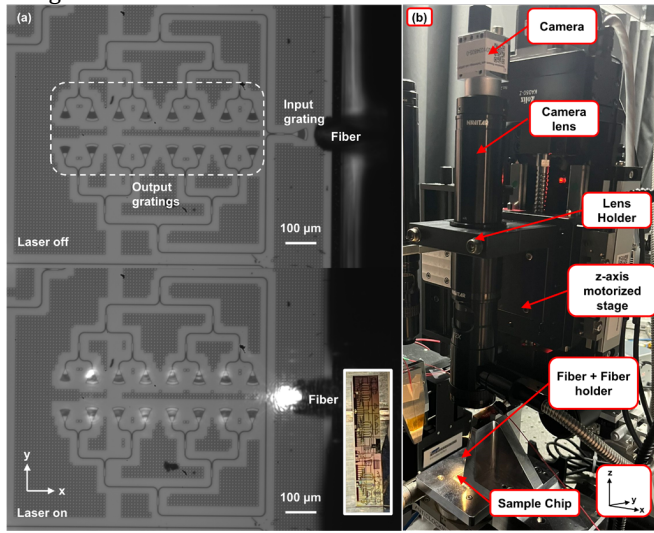

Fig. 2. (a) Microscopic images of SiPh gratings used in this work (inset: picture of the SiPh chip). (b) Camera system for real-time detection and auto focusing of light beams

From z = 0 µm to z = 800 µm, images taken in z = 0 – 250 µm, 250 – 400 µm, 400 – 600 µm, and 600 – 800 µm are categorized as Region A, Region B, Region C, and Region D, respectively, as illustrated in Fig. 1. The fundamental goals of this work are to develop a YOLO-based system which can detect light beams in Region A, B, C, and D in real-time; and automatically shift the camera lens to Region B to obtain focused light beams when beams in Region A, C, or D are detected. For the preparation of YOLO model training, the light beams on each image are labelled in bounding boxes, in the form of "<class_id> (which region is the beam located?), <x_center>, <y_center>, <width>, <height> (normalized geometrical indicators of the bounding box)". Fig. 3 illustrates the typical labelling of light beams in a captured image. The mathematical concepts behind YOLO model are explained in ref. [14], while the details of YOLO model training in Python platform are shown in the full Python code in ref. [15]. After labelling all 3,200 images, the images are split 50/50 into training datasets and validation datasets for model training. The training is carried out with 200 epoch, with several data augmentations implemented, as illustrated in the full Python code given in ref. [15].

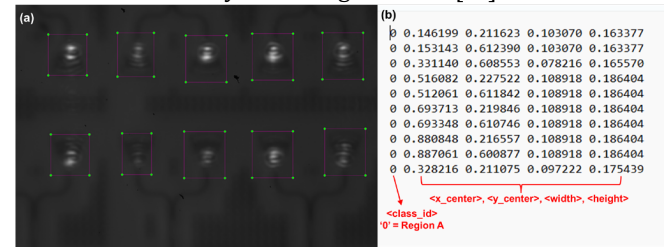

Fig. 3. Example of image labelling: (a) Light beams with bounding boxes, (b) labeling of each bounding box

Two fundamental losses from the training and validation datasets are obtained: bounding box losses and classification losses. Bounding box losses measure how well the predicted bounding boxes match the ground truth (labeled light beam), while the classification losses measure how well the model predicts the correct regions for the light beam inside the bounding boxes. Fig. 4 shows the training and validation losses for both bounding boxes and classifications. At 200th epoch, the training and validation losses for both bounding boxes and classifications reduces, and the curves for both training and validation losses coincides. This indicates that the trained model has minimal underfitting and overfitting [16].

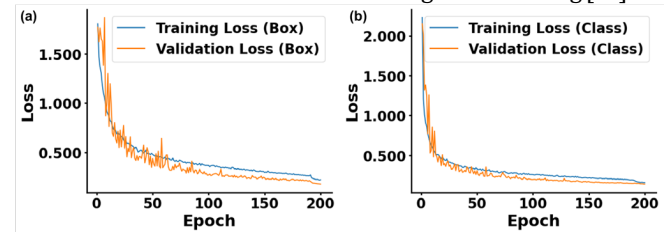

Fig. 4. Training and validation losses of: (a) bounding boxes, (b) classifications

**YOLO Model Testing.** The trained YOLO model is used to predict the images in the validation datasets, which was not fed into the YOLO model during the training process. From Fig. 5, the morphologies of light beam in different regions can be observed. In Region A, the light beam is unfocused. At the same time, the grating structures are still visible. In Region B, the grating structures are not visible, with focused light beams coupled out from the gratings. In Region C, the light beams are slightly distorted. In Region D, the light beams are highly sparse. In Fig. 5, all light beams in different regions are detected, with high confidence level of >0.95. This indicates that the YOLO model can accurately detect the light beams at their corresponding regions.

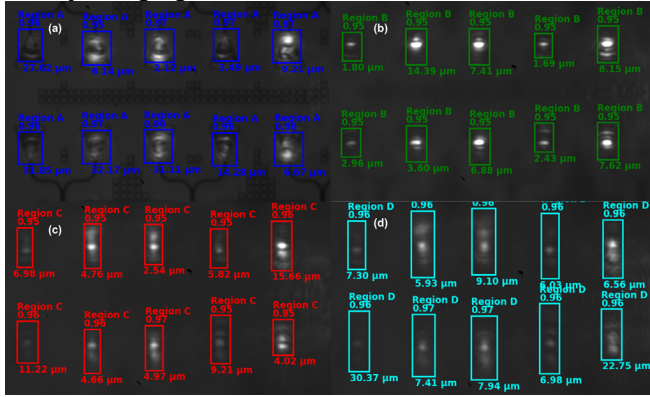

Fig. 5. Detected beams from validation datasets taken in: (a) Region A, (b) Region B, (c) Region C, and (d) Region D

The confusion matrix when using the trained YOLO model to detect light beams in 1,600 validation images is shown in Fig. 6(a). Within Region A, B, C and D, the YOLO model correctly classified 99.7% of the detected beams. This outstanding result enable us to use the trained YOLO model for real-time detection, which will be discussed later. However, in the last column, there are instances where the YOLO model mistakenly thought there was a light beam, but it might be the light scattering or noises in the camera. Meanwhile, the boxplots in Fig. 6(b) shows that most detected beams have confidence levels of >0.95. However, there are significant number of outliers where low confidence levels are obtained, ranging from 0.2 to 0.8.

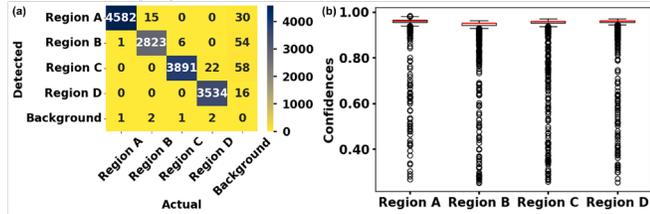

Fig. 6. (a) Confusion matrix obtained from detecting the validation images using trained YOLO model, (b) Boxplots of the confidence levels in Region A, B, C, and D

The outliers of confidence levels in Fig. 6(b) may occur at the boundaries between Region A, B, C, and D. For further investigation, we used the trained YOLO to detect light beams in images at z = 249 µm and z = 400 µm, as shown in Fig. 7. It was observed that there are overlapping features of Region A/B and Region B/C in the detected beams. Generally, the overlapping features have low confidence level of <0.5, which explains the occurrence of outliers in Fig. 6(b). These overlapping features should be taken into consideration when designing the software architecture of the auto-focusing system integrated in the real-time detection.

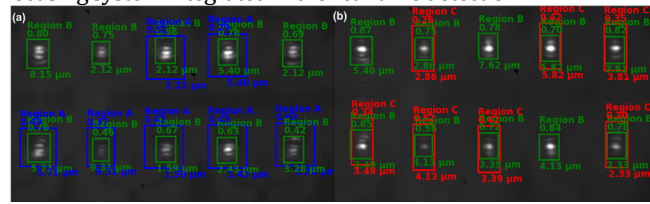

Fig. 7. Detected beams from validation datasets taken in: (a) z = 249 µm, (b) z = 400 µm

**Auto-focusing System.** After obtaining satisfactory outcomes from testing the trained YOLO model, the model is implemented into the auto-focusing system. The software architecture of the auto-focusing system is illustrated in Fig. 8. First, the CMOS camera shown in Fig. 2(b) will continuously capture images of the light beams from the SiPh sample chip at~17 frame-per-second (fps). At each frame, several light beams are detected, and the detected classes (Region A, B, C, or D) are stored in a list. If >50% of the detected beams are in Region A, the camera moves up continuously by moving the motorized z-axis stage, until a point where less than 50% of the detected beams are in Region A. However, due to possible overlapping features shown in Fig. 7, we included an additional condition where, as long as <80% of the detected beams are not in Region B, the camera will continue to move up until >80% of the detected beams are focused Region B beams. Similar mechanisms are applied to Region C and D. If >50% of the detected beams are in Region C or D, the camera moves down continuously until >80% of the detected beams are focused Region B beams. The full Python code for Fig. 8 is given in ref. [17]. The real-time detection and auto-focusing features of the developed system is demonstrated in a video uploaded to YouTube[12]. For real-time detection of beams in Region A, B, C, and D, the demonstration starts from 0:00 of the video. For auto-focusing from Region A to Region B, the demonstration starts from 0:34 of the video. For auto-focusing from Region D to Region B, the demonstration starts from 2:42 of the video.

The auto-focusing system developed in this work includes three key features, real-time detection of light beams, computation of beam widths of the detected beams, and auto-focusing of light beams from Region A/C/D to Region B. For the computation of beam widths, the computation is carried out by converting the images cropped by the detected bounding boxes to gray-scale images with pixel values ranging from 0 – 255. Thus, it provides a rapid solution in determining the beam width of light coupled out from SiPh gratings at focused region (Region B). Further expansion of this work is possible. The code can be modified to let user specify a beam width. When the camera reaches a point where >80% of the detected beams are in Region B, the camera moves up and down continuously till a point where the specified beam width is obtained. To facilitate further expansion of this work, we have included the full Python code used for YOLO model training, real-time beam detection, and auto-focusing in ref. [15,17].

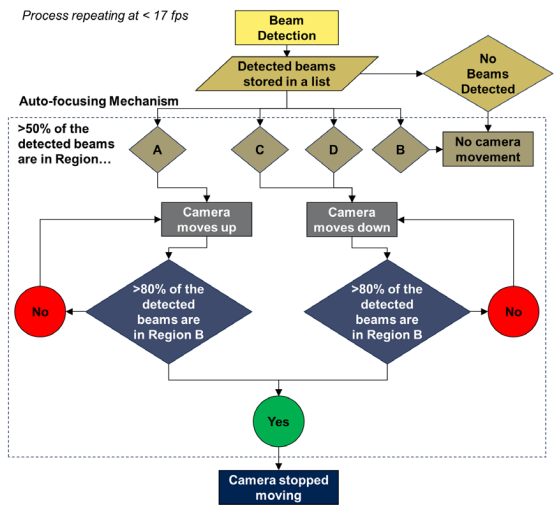

Fig. 8. Software architecture of the auto-focusing system

In summary, we developed a YOLO-based auto-focusing system to obtain focused profiles of light coupled out from silicon photonics (SiPh) gratings. The trained YOLO model shows classification accuracy of 0.997, and the confidence levels of the detected bounding boxes are mostly above 0.95. We demonstrated the real-time detection and auto-focusing features of the developed system in the attached video uploaded to YouTube.

**Back Matter**

**Funding.** Ministry of Education of Singapore AcRF Tier 2 (T2EP50121-0002 (MOE-000180-01)) and AcRF Tier 1 (RG135/23, RT3/23); National Semiconductor Translation and Innovation Centre (NSTIC (M24W1NS007));  National Centre for Advanced Integrated Photonics (NCAIP) (NRF-MSG-2023-0002); National Research Foundation, Singapore, and A*STAR under its Quantum Engineering Program (NRF2021-QEP2-03-P07) and A*STAR SPF (C222517002).

**Acknowledgment.** This work was supported by the Ministry of Education of Singapore AcRF Tier 2 (T2EP50121-0002 (MOE-000180-01)) and AcRF Tier 1 (RG135/23, RT3/23); National Semiconductor Translation and Innovation Centre (NSTIC (M24W1NS007));  National Centre for Advanced Integrated Photonics (NCAIP) (NRF-MSG-2023-0002); National Research Foundation, Singapore, and A*STAR under its Quantum Engineering Program (NRF2021-QEP2-03-P07) and A*STAR SPF (C222517002). The preparation of Python codes in ref. [15,17] is partly assisted by the generative AI tool, ChatGPT.

**Disclosures**. The authors declare no conflicts of interest.

**Data Availability Statement (DAS).** Full Python code is given in ref. [15,17]. Data, including the training/validation losses and training/validation images, are available upon requestion.